\documentclass{aastex631}

\shorttitle{Weak energy releases in the `quiet' Sun}
\shortauthors{Ramesh et al.}

\usepackage{hyperref}
\hypersetup{linkcolor=red,citecolor=blue,filecolor=cyan,urlcolor=magenta}

\usepackage{amsmath}
\usepackage{diagbox}

\usepackage{}

\begin{document}

\title{Radio, X-ray and extreme-ultraviolet observations of weak energy releases in the `quiet' Sun}

\author{R. Ramesh}
\affiliation{Indian Institute of Astrophysics, Koramangala 2nd Block, Bangalore, Karnataka, India - 560034}

\author{C. Kathiravan}
\affiliation{Indian Institute of Astrophysics, Koramangala 2nd Block, Bangalore, Karnataka, India - 560034}

\author{N. P. S. Mithun}
\affiliation{Physical Research Laboratory, Navrangpura, Ahmedabad, Gujarat, India - 380009}

\author{S. V. Vadawale}
\affiliation{Physical Research Laboratory, Navrangpura, Ahmedabad, Gujarat, India - 380009}

\begin{abstract}

We analyzed ground-based low frequency ($<$100\,MHz) radio spectral and imaging data of the solar corona obtained with the facilities in the Gauribidanur observatory during the same time as the very weak soft X-ray flares
(sub A-class, flux $\rm {<}10^{-7}\,Wm^{-2}$ in the 1\,-\,8\,$\rm {\AA}$ wavelength range) from the `quiet' Sun observed with the X-ray Solar Monitor (XSM) onboard Chandrayaan-2 during the recent solar minimum. Non-thermal type I radio burst activity were noticed in close temporal association with the X-ray events. The estimated brightness temperature ($T_{b}$) of the bursts at a typical frequency like 80\,MHz is ${\approx}3{\times}10^{5}$\,K. Extreme-ultraviolet (EUV) observations at 94{\AA} with the Atmospheric Imaging Assembly (AIA) onboard the Solar Dynamics Observatory (SDO) revealed a brightening close to the same location and time as the type I radio bursts. As far as we know reports of simultaneous observations of 
X-ray and/or EUV counterpart to weak transient radio emission at low frequencies from the `quiet' Sun in particular are rare. Considering this and the fact that low frequency radio observations are sensitive to weak energy releases in the solar atmosphere, the results indicate that coordinated observations of similar events would be useful to understand transient activities in the `quiet' Sun. 
\end{abstract}


\section{Introduction} \label{sec:intro}

A variety of small scale energy releases on the Sun (flaring bright points, active region transient brightenings, coronal jets, etc.) have been studied using X-ray and radio observations. While X-rays are dominated by thermal emission from the coronal plasma, radio observations are sensitive to non-thermal emission also. The observations of low frequency type III radio bursts in association with X-ray bright point flares \citep{Kundu1980,Kundu1994} clearly indicated that the latter are capable of accelerating particles to non-thermal energies, as well as producing the heated material detected in soft X-rays. 
The detection of type III bursts together with coronal X-ray jets strengthened the above argument \citep{Aurass1994,Kundu1995}. 
These results imply that radio observations are an useful complimentary tool for observing signatures of weak, transient energy releases in the solar atmosphere since the related non-thermal emission can be easily detected \citep{Benz1995,Mugundhan2017}. 
Note that counterparts to some of the X-ray transients have been reported at higher radio frequencies also.  
For e.g., \cite{Gopalswamy1994,White1995,Gary1997} observed correlated active region transient brightenings in soft X-rays and microwaves. Moving further,
X-ray microflares are another independent observational evidence for the small scale energy releases in the solar atmosphere. They were first reported by \cite{Lin1984}.  The energy involved (${\sim}10^{26}$\,erg) is approximately six orders of magnitude lower than the corresponding value for some of the largest solar flares. 
Sensitive observations with the soft X-ray telescope onboard YOHKOH revealed that the microflares are present in the `quiet' Sun also \citep{Krucker1997}. 
The study of these microflares are of interest because of their possible bearing on the problems of coronal heating and solar flares \citep{Hudson1991,Hannah2011,Benz2017}. Analogous to microflares, \cite{Kundu1986} reported observations of weak non-thermal microbursts in the solar corona at low radio frequencies. 
Though it was hinted that a common source of energetic particles could be responsible for both the microflares and microbursts, reports of direct association are rare. The microbursts were found to have some characteristics similar to that of the normal type III bursts, but the relationship was inconclusive. Further the  observations reported were at separate individual frequencies unlike typical spectral observations of type III bursts \citep{Kundu1986,White1986,Thejappa1990,Subramanian1993}. 
Recent spectroscopic imaging observations indicate that the weak non-thermal radio emission at low frequencies are more like type I radio bursts \citep{Sharma2018,Mondal2020}.
However there were no details about the counterparts to the radio events in other frequency bands of the electromagnetic spectrum. Note that type I bursts represent the smallest discrete releases of energy observable \citep{Bastian1998}. They are considered to be evidence of successive electron accelerations. So, establishing its association with activities in other regions of the solar atmosphere would be useful to understand the acceleration processes of the non-thermal electrons at the sites of elementary/weakest energy releases. 
In this situation, we report observations of weak type I radio burst emission during the same time as soft X-ray observations of a sub-A class level flare and EUV brightening from the  `quiet' solar corona in the complete absence of active regions and flare/coronal mass ejection (CME) activity. 

\section{Observations} \label{sec:obs}

The radio observations were carried out using the different facilities operated by the Indian Institute of Astrophysics (IIA) in the Gauribidanur Observatory\footnote{\url{https://www.iiap.res.in/?q=centers/radio}} \citep{Ramesh2011a,Ramesh2014}. The radio spectral images were obtained with the
Gauribidanur LOw-frequency Solar Spectrograph (GLOSS) in the frequency range 85\,-\,40 MHz \citep{Ebenezer2001,Ebenezer2007,Kishore2014,Hariharan2016b}. The GLOSS is an one-dimensional array of eight log-periodic dipole antennas (LPDA) along a North-South baseline. The half-power width of the response pattern of GLOSS around local noon is 
${\approx}90{\arcdeg}{\times}6{\arcdeg}$ (right ascension, R.A.\,{$\times$}\,declination, decl.) at the highest frequency of operation, i.e. 85\,MHz. While the width of the response pattern along R.A. is nearly independent of frequency, its width along the declination varies inversly with the frequency due to  interferometric arrangement of the individual antennas. The observations were carried out with an integration time of ${\approx}$1\,sec and bandwidth of 
${\approx}$1\,MHz. The minimum detectable flux density is ${\approx}$75\,Jy 
(1\,Jy\,=\,$\rm 10^{-26}\,Wm^{-2}Hz^{-1}$) at a typical frequency like 80 MHz.
The antenna and the receiver systems were calibrated by carrying
out observations in the direction of the Galactic center as described in \cite{Kishore2015}. The two-dimensional radio images were obtained with the Gauribidanur RAdioheliograPH (GRAPH) at 80 MHz \citep{Ramesh1998,Ramesh1999a,Ramesh2006b}. The GRAPH is a T-shaped radio interferometer array of 384 LPDAs. Its angular resolution (`beam' size) for observations close to the zenith is 
${\approx}5{\arcmin}{\times}7{\arcmin}$ (R.A.\,$\times$\,decl.) at the above frequency. The integration time is $\approx$250\,msec and the observing bandwidth is 
$\approx$2\,MHz. The field-of-view (FOV) in the GRAPH images is 
${\approx}2{\arcdeg}{\times}2{\arcdeg}$, and the pixel size is 
${\approx}14{\arcsec}$. The minimum detectable flux density is 
${\approx}$2\,Jy. The GRAPH data were calibrated using the standard Astronomical Image Processing System (AIPS). The combined use of the imaging and spectral data help to understand the radio signatures associated with the corresponding solar activity in a better manner (see for e.g. \cite{Sasikumar2014}).

\begin{figure}[t!]
\centerline{\includegraphics[angle=0,height=7.5cm,width=12cm]{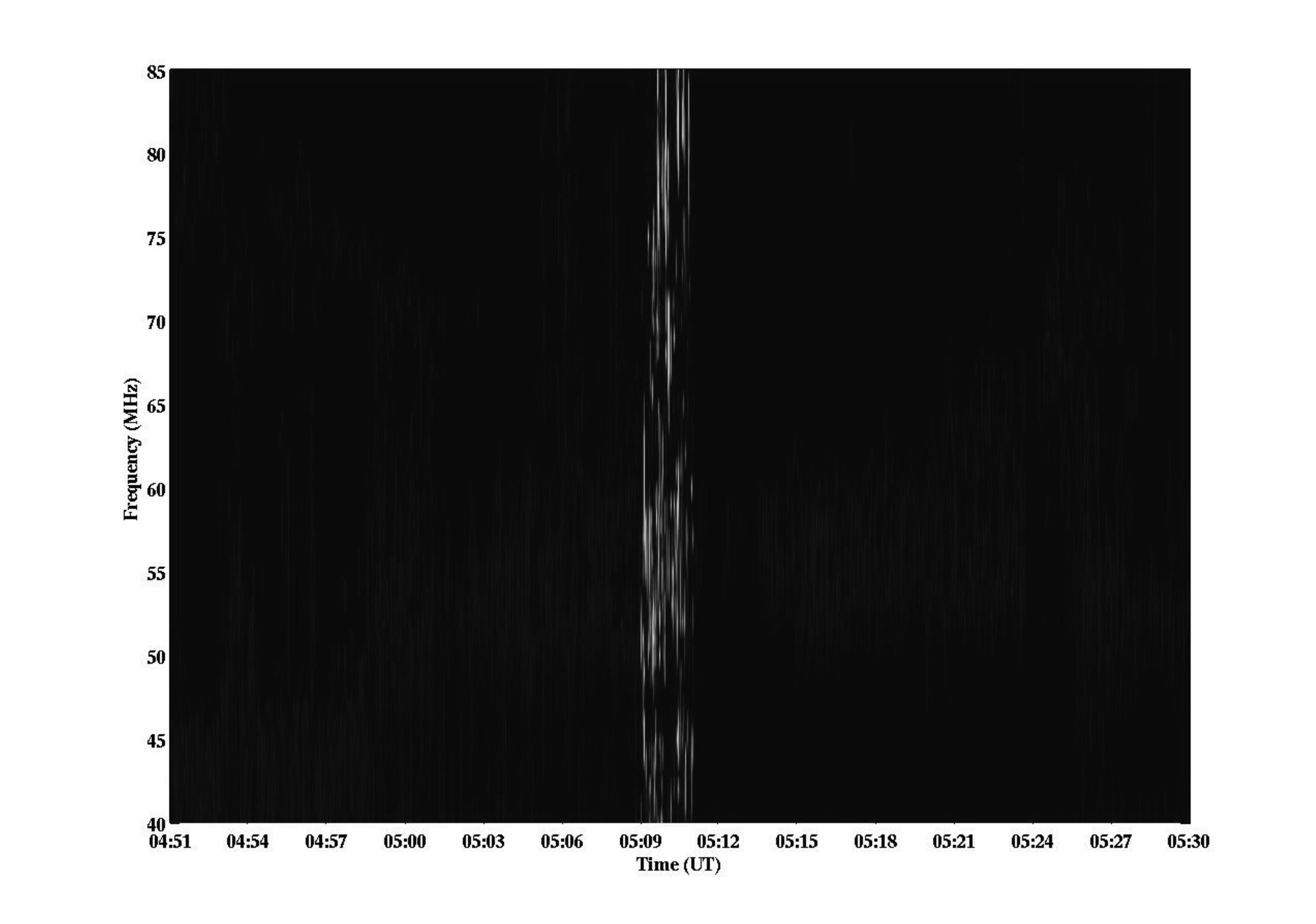}}
\caption{GLOSS dynamic spectrum of the solar radio emission observed on 2020 April 21. The bright emission during the period $\approx$05:09\,-\,05:11\,UT correspond to the type I solar radio bursts mentioned in the text.}
\label{fig:figure1}
\end{figure}

\begin{figure}[t!]
\centerline{\includegraphics[angle=0,height=7.5cm,width=12cm]{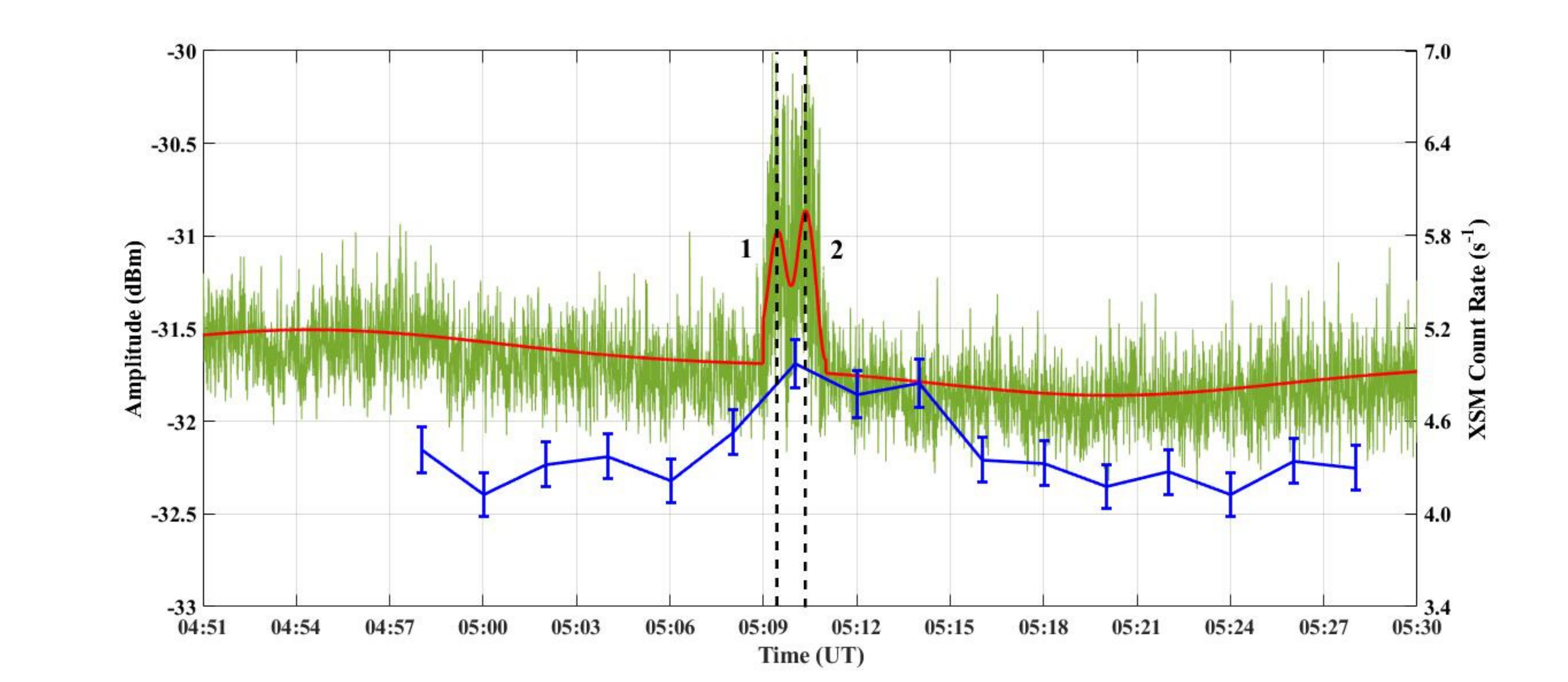}}
\caption{The `green' colour plot corresponds to the frequency averaged time profile of the GLOSS dynamic spectrum in Figure \ref{fig:figure1}. Its amplitude values are indicated in the left hand side ordinate axis. The `red' colour line is the fit to the data points. The labels 1 \& 2 indicate the epochs of maximum radio emission from the regions 1 \& 2 in Figure \ref{fig:figure3}, respectively. The `blue' colour profile is the light curve of the soft X-ray emission from the Sun close to the same epoch as the radio observations. Its amplitude values are indicated in the right hand side ordinate axis. The data were obtained with the Chandrayaan-2/XSM \citep{Mithun2020} in the energy range 
$\approx$1\,-\,5\,keV with a time binning of ${\approx}$120\,sec.}
\label{fig:figure2}
\end{figure}

\begin{figure}[t!]
\centerline{\includegraphics[angle=0,height=6cm,width=12.5cm]{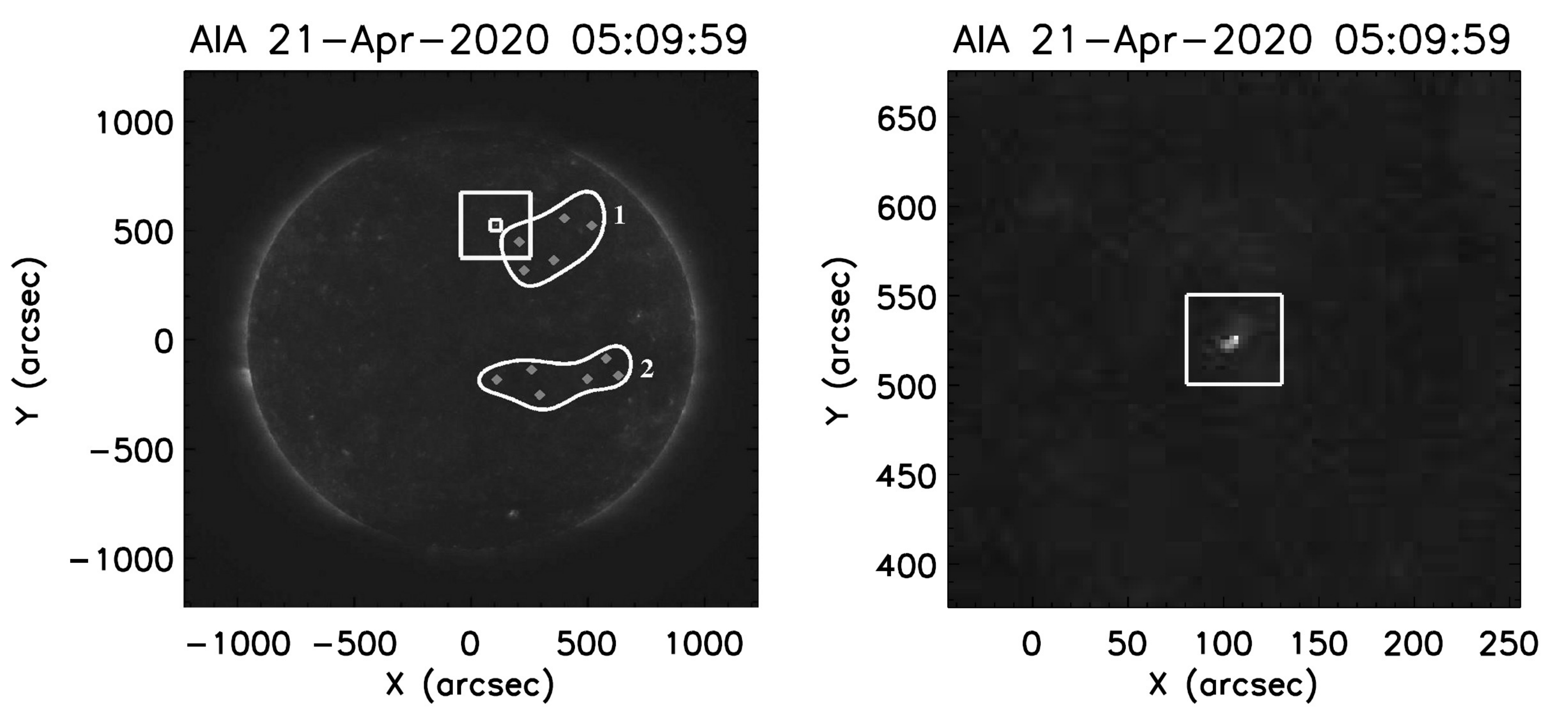}}
\caption{A composite of the GRAPH difference image of the bursts in Figure \ref{fig:figure1} at 80 MHz and the EUV observations at 94{\AA} with the SDO/AIA \citep{Lemen2012} around the same time as the radio and X-ray observations in Figures \ref{fig:figure1} \& \ref{fig:figure2} on 2020 April 21. The contours labelled 1 \& 2 correspond to the GRAPH observations. The background is the EUV image. Solar north is straight up and east is to the left. The bigger and smaller `boxes' in the left panel image indicate the region around the EUV brightening and the location of maximum emission, respectively. The `zoomed' version of the same brightening is shown in the right side panel. The peak flux density in the GRAPH observations is $\rm {\approx}\,241\,Jy$. Its nearly the same for the contours 1 \& 2, which correspond to the two maxima 1 \& 2 in the radio time profile in Figure \ref{fig:figure2}, respectively. The contours shown are at 80\% level. The `dots' inside the contours 1 \& 2 indicate the centroid locations of the individual type I bursts a\,-\,e \& f\,-\,k in Figure \ref{fig:figure4}, respectively.}
\label{fig:figure3}
\end{figure}

Figure \ref{fig:figure1} shows the GLOSS observations on 2020 April 21 in the time interval 04:51\,-\,05:30\,UT. The patches of bright emission during the period 
$\approx$05:09\,-\,05:11\,UT are typical of  type I or noise storm bursts from the solar corona (see for e.g. \cite{Iwai2013,Mugundhan2018b}). It is widely believed that the bursts are due to plasma radiation at the fundamental plasma frequency \citep{Melrose1980,Kai1985}.
Figure \ref{fig:figure2} shows the frequency averaged time profile of the dynamic spectrum in Figure \ref{fig:figure1}. The presence of enhanced activity during the interval $\approx$05:09\,-\,05:11\,UT could be clearly noticed. It is also similar to the time profiles of groups of type I bursts (see for e.g. \cite{Ramesh2013b,Mugundhan2016}).
No H$\alpha$ and/or GOES soft X-ray flares were reported during the burst interval mentioned above\footnote{\url{https://www.solarmonitor.org/data/2020/04/21/meta/noaa{\_}events{\_}raw{\_}20200421.txt}}. The Sun was totally free of any active regions\footnote{\url{https://www.solarmonitor.org/?date=20200421}} and/or CMEs\footnote{\url{https://cdaw.gsfc.nasa.gov/CME{\_}list/UNIVERSAL/2020{\_}04/univ2020{\_}04.html}}. 
The overall location of the bursts can be inferred from the GRAPH difference image (obtained by subtracting a pre-event image to clearly identify the weak emission features) in Figure \ref{fig:figure3} at 80\,MHz. 
The two spatially separated contours marked 1 \& 2 correspond to the two maxima (indicated by the same set of numbers) in the time profile of the bursts in Figure \ref{fig:figure2}.
The brightness temperature ($T_{b}$) of the contours 1 \& 2, estimated using the `beam' size of the GRAPH at 80\,MHz, are ${\approx}3{\times}10^{5}$\,K. The `dots' inside the contours in Figure \ref{fig:figure3} correspond to the centroids of some of the individual type I bursts (see Figure \ref{fig:figure4}). We located them following the methodology described in \cite{Ramesh2020a}.
Any ionospheric refraction effects on the radio source positions in the present case are expected to be very minimal since the observations were carried out close to the local noon during which time the zenith angle of the Sun is the least. 
Note that the elevation of Sun on 2020 April 21 during the present radio observations was ${\approx}90^{\arcdeg}$. 
Secondly, the total duration of the type I radio bursts in the present case is only 
$\approx$2\,min (see Figures \ref{fig:figure2} \& \ref{fig:figure4}). This is less than the period 
of $\approx$20\,min over which radio source positions at low frequencies usually change due to ionospheric effects \citep{Stewart1982,Mercier1996}. 

\section{Analysis and Results} \label{sec:anares}

Recently \cite{Vadawale2021} had reported observations of `quiet' Sun X-ray microflares with the Chandrayaan-2/XSM during the solar minimum 2019-2020 (see for e.g. \cite{Ramesh2020b}). Upon inspection we found that some of these flares were observed during the same epoch as the low frequency radio observations of the Sun from Gauribidanur. We considered the X-ray flare observed on 2020 April 21 at 
${\approx}$05:10 UT (see Figure \ref{fig:figure2}) for the present work since both radio spectral and imaging observations were available. There was also an EUV brightening observed with the
SDO/AIA at 94{\AA} (see Figure \ref{fig:figure3}) around the same time as the type I radio bursts (Figures \ref{fig:figure1} \& \ref{fig:figure2}) and the X-ray flare (Figure \ref{fig:figure2}). The location of the northern radio contour with label `1' in Figure \ref{fig:figure3} correspond reasonably well with the location of the EUV brightening. The observations of the type I radio bursts over a larger area compared to the EUV brightening could be due to the divergence of the associated field lines (see for e.g. \cite{Li2017}). We speculate that the presence of the two spatially separated radio contours 1 \& 2 (particularly with the latter located just below the equator in the southern hemisphere) suggests interaction at two different locations between inclined, large magnetic loops with foot points in the same hemisphere, north in the present case \citep{Wild1968,Simnett1998}. Note that the probability of trans-equatorial loops are expected to be minimal since there were no active regions on the solar disk. Information on the polarization characteristics of the regions 1 \& 2 would have helped to verify the above. But, observations with the GRAPH in its current configuration are limited to the total intensity mode.
We also checked the location of the 1st sidelobe in the GRAPH `beam', particularly in the north-south direction, to rule out the possibility of any spurious pick-up. It was $\gtrsim\,14^{\arcmin}$ away from the main lobe. The spacing between the contours 1 \& 2 in Figure \ref{fig:figure3} is shorter compared to this. Secondly, the amplitude of the sidelobe is lesser by a factor of 
20 (${\approx}$\,13\,dB). But the strength of the two sources 1 \& 2 are nearly the
same.

The peak flux of the XSM flare is 
$\rm {\approx}6{\times}10^{-9}\,Wm^{-2}$. It was a very weak event (see 
Figure \ref{fig:figure2}).  
The total duration of the event is ${\approx}$5\,min. There appears to be two  `peaks' in the flare light curve with a noticeable difference between the corresponding count rates.
The type I radio bursts are present only during the initial phase of the X-ray emission, i.e. close to the 1st of the two `peaks' mentioned above.
The total duration of the radio event is smaller $({\approx}$2\,min).  
Assuming that both the X-ray and radio events are related to a common primary phenomenon, the comparitively shorter duration of the radio event indicates that  the electrons responsible for its occurrence are probably thermalized quickly. As a result they cannot travel to larger heights in the corona from where the low frequency radio emission primarily originate \citep{Mondal2020}. The shorter duration of the radio bursts could be also due to the emission being non-thermal in nature as compared to the soft X-ray emission \citep{Reid2017}. Nevertheless we independently calculated the associated energy from the radio observations. 

\begin{figure}[t!]
\centerline{\includegraphics[angle=0,height=7.5cm,width=12cm]{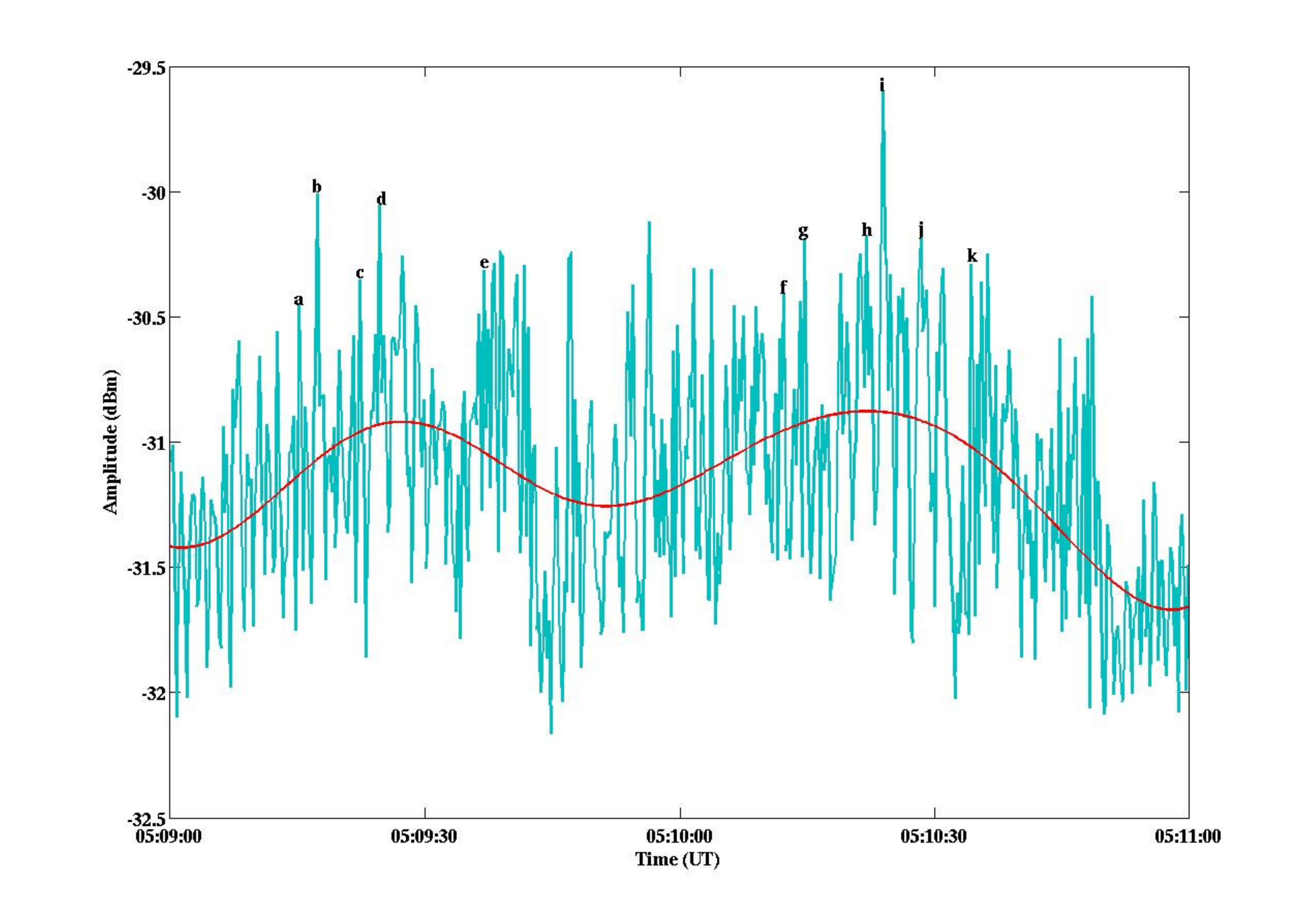}}
\caption{`Zoomed' version of the observations around the maxima in the radio time profile in Figure \ref{fig:figure2}. The labels a\,-\,k indicate some of the individual type I bursts.}
\label{fig:figure4}
\end{figure}

The energy associated with a type I burst can be estimated using the relation $\rm E\,{=}\,S{\delta}t{\delta}{\nu}R^{2}{\Omega}e^{\tau}$ \citep{Elgaroy1977}. Here $\rm S$ is the flux density of the burst, $\rm {\delta}$t is the duration of the burst, ${\nu}$ is the frequency of observation, $\rm {\delta}{\nu}$ is the bandwidth of the burst, $\rm R$ is the Sun-Earth distance, ${\Omega}$ is the solid angle into which the radio waves are emitted, and $\rm \tau$ is the optical depth. In the present case, $\rm S\,{\approx}\,241\,Jy$ (see Figure \ref{fig:figure3}), 
$\rm {\delta}t\,{\approx}\,$1\,sec, and $\rm {\delta}{\nu}\,{\approx}\,5\,MHz$ near 80\,MHz (see Figure \ref{fig:figure1}). Assuming 
${\Omega}\,{=}\,$0.15\,steradians \citep{Steinberg1974}, $\rm {\tau}\,{\approx}$\,3 at 80 MHz \citep{Ramesh2005b}, and an efficiency ($\eta$) of ${\approx}10^{-10}$ for the type I burst emission process \citep{Prasad2004}, we find $\rm E\,{\approx}\,8.1{\times}10^{22}$\,erg. This is consistent with the reports that ${\sim}10^{21}$\,{-}\,$10^{23}$\,ergs are needed for a single type I burst \citep{Tomin2018}. 
We also calculated the energy using the relation $\rm E\,{=}\,n_{th}(n/n_{th})VE_{m}$ (see for e.g. \cite{Ramesh2010c}). Here $\rm n_{th}$ is the number density of the background thermal electrons, $\rm n$ is the number density of the non-thermal electrons, $\rm V$ is the volume of the burst source, and $\rm E_{m}$ is the mean energy of the individual electrons. In the present case $\rm n_{th}\,{=}\,7.9{\times}10^{7}\rm cm^{-3}$ and $\rm E_{m}\,{\approx}\,5$\,keV \citep{Vadawale2021}. Assuming $\rm n/n_{th}\,{=}\,1.23{\times}10^{-7}$ at 80 MHz \citep{Thejappa1991} and 
V\,=\,$10^{30}\rm cm^{3}$ (corresponding to a density scale height of 
$\rm {\approx}\,10^{10}\,cm$ in the solar corona), we find $\rm E\,{\approx}\,7.8{\times}10^{22}$\,erg. This is in good agreement with the estimated energy using the observed flux density, duration, bandwidth, etc. of the burst  in the present case.
We would like to mention here that the noise storm radiative efficiency $\eta$ mentioned above is typically in the range ${\sim}\,10^{-6}$\,-\,$10^{-10}$. In the present case both the type I bursts and the associated X-ray emission were short lived. Therefore it is likely that the electron acceleration responsible for the type I bursts were triggered by the same process responsible for the associated X-ray microflare \citep{Crosby1996}. The energy of the latter is typically ${\sim}\,10^{27}$\,erg. 
Reports indicate that for such an energy input, $\eta$ is expected to be in the range ${\sim}\,10^{-9}$\,-\,$10^{-10}$ \citep{Prasad2004}. We assumed $\eta\,{\approx}\,10^{-10}$ since the observed type I bursts were also weak. The close agreement between the different energy estimates mentioned above is in support of our assumption on the value of $\eta$. However it should be kept in mind that the above calculations will give rise to a lower energy for the type I burst if we assume $\eta\,>\,10^{-10}$. Hence a more tighter constraint on the value of $\eta$ would be better.

Proceeding further, we find that the area enclosed by the contours in Figure \ref{fig:figure3} is nearly same as that of the GRAPH `beam' size at 80 MHz mentioned earlier, i.e.\,${\approx}5^{\arcmin}{\times}7^{\arcmin}$. But results obtained from (i) high angular resolution observations of the solar corona at low radio frequencies during solar eclipses (lunar occultation technique) and (ii) independent long baseline interferometer observations indicate that the `true' size of the individual type I bursts is ${\lesssim}15^{\arcsec}$ \citep{Ramesh2001b,Kathiravan2011,Ramesh2012b,Mugundhan2016,Mugundhan2018a}. There are also reports that the upper limit to the size of a type I burst source is 
${\approx}14^{\arcsec}$ \citep{Melrose1980}. These values are much smaller than the GRAPH `beam' size. \cite{Kundu1990,Malik1996,Willson1997} had shown earlier that the centroids of type I burst sources are spatially distributed within the associated noise storm emitting region. 
Type I burst models too predict scattered small-scale sites of energy release \citep{Klein1995}.
The dispersion in the centroids of some of the individual type I bursts in the present case (see Figures \ref{fig:figure3} \& \ref{fig:figure4}) is consistent with this. 
Therefore it is possible that the contours in Figure \ref{fig:figure3} correspond to an ensemble of type I burst sources, each of size ${\approx}14^{\arcsec}{\times}14^{\arcsec}$. So, we calculated the maximum possible total energy of the type I bursts as   
$\rm E_{t}\,{\approx}\,\frac{5{\times}7{\times}3600{\times}8.1{\times}10^{22}}{14{\times}14}\,{\approx}\,5.3{\times}10^{25}$\,erg. 
This is in reasonable agreement with the range of energies 
($3{\times}10^{26}$\,-\,$7{\times}10^{27}$\,erg) for the soft X-ray microflares reported by \cite{Vadawale2021} since the authors had mentioned that their estimates represent upper limits. Note that the minimum possible energy of the type I bursts in the present case is $\rm E\,{\approx}\,8.1{\times}10^{22}$\,erg. So, our estimates indicate a range of ${\approx}10^{22}$\,-\,$10^{25}$ erg.
\cite{Benz2002} had earlier reported EUV flares in the `quiet' solar corona with energy budget ${\approx}10^{24}$\,-\,$10^{26}$ erg. 
\cite{Lin1985} showed that the total energy released into the interplanetary medium in  solar electrons above 2\,keV is ${\approx}10^{25}$\,-\,$10^{26}$ erg. 
The above numbers and arguments confirm that the type I radio bursts are an independent ground based observational tool to probe weak activity in the `quiet' regions of the corona also in addition to its known association with sunspot activity (see for e.g. \cite{Ramesh2000b}).

\section{Summary} \label{sum}

We had presented co-temporal/co-spatial observations of weak type I radio bursts, X-ray microflare, and EUV brightening from the `quiet' Sun which was completely devoid of any active regions. There is close agreement between the energy budgets estimated independently from the radio and X-ray observations. As far as we know, this is the first time such simultaneous observations of transient activity in the `quiet' Sun have been reported. Considering that type I radio bursts like those described in this work hint activity in the outer layers of the solar corona which is currently inaccessible to observations in X-rays and extreme ultra-violet (EUV), combined investigations of weak energy releases observed at the same time in all the aforementioned domains would be helpful to understand the energies deposited at different levels in the solar corona in addition to the associated mechanisms themselves. For e.g. \cite{Li2017} had shown that magnetic reconnection driven by multiple moving magnetic features \citep{Harvey1973,Bentley2000} in/near an active region at the photosphere are correlated with EUV brightenings and type I bursts. But there were reports of 
H$\alpha$ and X-ray flares during the observing period reported by the above authors. Several active regions too were present. Nevertheless, it would be interesting to explore such moving features in the `quiet' Sun also to explain weak energy releases as described in this work. 

We express our gratitude to the Gauribidanur observatory staff members for their help in the observations and upkeep of the facilities. M.Rajesh and K.P.Santosh are acknowledged for their assistance to the present work. The SDO/AIA data are courtesy of the NASA/SDO and the AIA science teams. We thank the referee for his/her kind comments which helped us to describe the results more clearly.

\end{document}